\documentclass[twocolumn, superscriptaddress, prb, showpacs]{revtex4}

\usepackage{ifthen}
\usepackage{graphicx}
\usepackage{amsmath}

\newcommand{\rr}{{\bf r}}
\newcommand{\RR}{{\bf R}}
\newcommand{\rt}{{\bf\tilde{r}}}
\newcommand{\redgrad}{{\bf s}}
\newcommand{\kF}{k_{\text{F}}}
\newcommand{\dr}{d^3\mspace{-2mu}r}
\newcommand{\eq}[1]{Eq.~(\ref{#1})}
\newcommand{\eqs}[1]{Eqs.~(\ref{#1})}

\begin{document}

\title{Van der Waals density functional: Self-consistent potential and
the nature of the van der Waals bond}
\author{T. Thonhauser}
\author{Valentino R. Cooper}
\author{Shen Li}
\author{Aaron Puzder}
\affiliation{Department of Physics and Astronomy, Rutgers, The State 
University of New Jersey, Piscataway, New Jersey 08854-8019, USA.}
\author{Per Hyldgaard}
\affiliation{Department of Microtechnology and Nanoscience, Chalmers
University of Technology, SE-412 96 G\"oteborg, Sweden.}
\author{David C. Langreth}
\affiliation{Department of Physics and Astronomy, Rutgers, The State 
University of New Jersey, Piscataway, New Jersey 08854-8019, USA.}
\date{\today}

\begin{abstract}
We derive the exchange-correlation potential corresponding to the
nonlocal van der Waals density functional [M. Dion, H. Rydberg, E.
Schr\"oder, D. C. Langreth, and B.  I. Lundqvist, Phys.\ Rev.\
Lett.~{\bf 92}, 246401 (2004)]. We use this  potential for a
self-consistent calculation of the ground state properties of  a number
of van der Waals complexes as well as crystalline silicon. For the
latter, where little or no van der Waals interaction is expected, we
find that the results are mostly determined by semilocal exchange and
correlation as in standard generalized gradient approximations (GGA),
with the fully nonlocal term giving little effect. On the other hand,
our results for the van der Waals complexes show that the
self-consistency has little effect at equilibrium separations. This
finding validates previous calculations with the
same functional that treated the fully nonlocal term as a post GGA
perturbation. A comparison of our results with wave-function
calculations demonstrates the usefulness of our approach. The
exchange-correlation potential also allows us to calculate
Hellmann-Feynman forces, hence providing the means for efficient
geometry relaxations as well as unleashing the potential use of other
standard techniques that depend on the self-consistent charge
distribution. The nature of the van der Waals bond is discussed in terms
of the self-consistent bonding charge.
\end{abstract}

%%%%%%%%%%%%%%%%%%%%%%%%%%%%%%%%%%%%%%%%%%%%%%%%%%%%%%%%%%%%%%%%%%%%%%%%
% pac codes:
% 31.15.Ew    Density-functional theory 
% 71.15.Mb    Density functional theory, local density approximation,
%             gradient and other corrections
% 61.50.Lt    Crystal binding; cohesive energy
%%%%%%%%%%%%%%%%%%%%%%%%%%%%%%%%%%%%%%%%%%%%%%%%%%%%%%%%%%%%%%%%%%%%%%%%
\maketitle

%%%%%%%%%%%%%%%%%%%%%%%%%%%%%%%%%%%%%%%%%%%%%%%%%%%%%%%%%%%%%%%%%%%%%%%%
\section{Introduction}
%%%%%%%%%%%%%%%%%%%%%%%%%%%%%%%%%%%%%%%%%%%%%%%%%%%%%%%%%%%%%%%%%%%%%%%%

Density functional theory (DFT) has acquired high respect for its
simplicity and accuracy within first-principles calculations.  In
particular, generalized gradient
approximations\cite{perdew92,PBE,revPBE} have had much success in
describing isolated molecules\cite{GGAmolecule} as well as dense bulk
matter.\cite{GGAsolid} For molecules, this performance has been further
improved by the use of various empirical and hybrid methods. On the
other hand, for van der Waals (vdW) complexes and sparse matter,
including many layered structures, polymer crystals, and organic
molecular crystals, these methods either give sporadic results or fail
completely. The reason for this failure is that the dominant part of the
stabilization energy in many cases comes from the dispersion energy,
which is not correctly accounted for in standard 
DFT.\cite{spooner96,pulay94,ruiz95,rydberg03,dion04}

To remedy the situation, a new approach using a van der Waals density
functional (vdW-DF) with a nonlocal correlation energy has been
developed.\cite{dion04} This formalism includes van der Waals forces in
a seamless fashion and has been applied with quite good results to bulk
layered systems,\cite{rydberg03}
dimers,\cite{dion04,langrethIJQC,aaron05,Thonhauser06,svetla2,svetla1,
tsuzuki-naphthalene} molecules physisorbed on infinite
surfaces,\cite{benz-graph} and bulk crystals of polyethylene.\cite{pe}
Lacking the exchange-correlation potential corresponding to the vdW-DF,
these calculations evaluated the vdW-DF as a post-processing
perturbation, using a charge density obtained from self-consistent
calculations with the PBE or revPBE functionals.\cite{PBE,revPBE}

The goal of this paper is to derive the exchange-correlation potential
for the vdW-DF, thereby enabling fully self-consistent calculations using
vdW-DF. This exchange-correlation potential will give more credibility to
the post-process vdW-DF calculations mentioned above. Furthermore, it
encourages us to
apply the vdW-DF to even more exciting cases in the future. In addition,
the knowledge and implementation of the exchange-correlation potential
provides the missing underlying framework for further developments, such
as the evaluation of forces and the stress tensor. These tools are of
tremendous interest for the structural optimization of many soft
materials. Also many other standard tools for analyzing materials
properties that are usually built into ab-initio codes become accessible
through the self-consistent vdW-DF potential. With these
developments, vdW-DF should become a standard tool for systems where
vdW interactions are important. This includes not only extended systems,
but also finite
systems that are too large for the standard quantum chemical methods to
be effective.

For such large finite systems, the advantage of the vdW-DF approach
becomes obvious when the scaling of the computational effort with
respect to system size is considered. Particular interest in van der
Waals complexes has emerged in the context of large systems, where
questions regarding DNA base-pair bonding, protein structure and
folding, and organic molecule crystallization are experiencing a surge
of interest. Traditionally, second-order M{\o}ller--Plesset perturbation
theory\cite{mp2} (MP2) and coupled-cluster calculations\cite{ccsdt} with
singles, doubles, and perturbative triple excitations (CCSD(T)) would be
used to study these systems as they are regarded as accurate
and reliable.  Unfortunately, in many of these
cases their application is limited to relatively small systems, since
they are too computationally demanding. Furthermore, the results are
basis set dependent and elaborate techniques are necessary to estimate
results at the complete basis set limit.  This is where the strength of the
vdW-DF approach lies: Since it scales with system size $N$ just like
a regular DFT calculation---usually $\mathcal{O}(N^3)$, complexes
currently out of the reach for MP2 and CCSD(T) calculations are easily
treatable.

For extended systems that require the consideration of long-range van
der Waals interactions, the advantage of vdW-DF is even more obvious. To
our knowledge there is simply no other first-principles method
available. The earlier post-process application of the method has
already given reasonable results for a polymer crystal\cite{pe} and
layered intercalates,\cite{zimbaras07} and
applications to molecular crystals are under way.  The results of the
this work suggest that the predictions will not be greatly affected by
the full self-consistency that now becomes possible.

The vdW-DF treats the nonlocal vdW interaction by
an approximate non-empirical method that is  not only correct in
principle,  but useful in practice. It differs from other methods
\cite{elstner01,scoles01,wu02,Hasegawa04,Grimme04,Zimmerli04,elstner98,sanchez97}
that treat the vdW interaction as occurring directly between the nuclei
via empirically determined potentials. As discussed later, the correct
source for the forces on the nuclei is electrostatic, and results from a
change in the charge densities of the fragments as the vdW bond is formed. This
effect is included correctly in vdW-DF.

We have organized this paper in the following way. In
Sec.~\ref{sec:derivative} we analytically derive the
exchange-correlation potential. In Sec.~\ref{sec:results} the potential
is then applied to several, very different physical systems. We compare
our results with other calculations such as MP2 and CCSD(T) and find
good agreement.  Our results show that calculating vdW-DF binding
energies in a post-process procedure, rather than self-consistent, is
reasonable, thus justifying the approximation used in earlier
calculations. Here, we also deal with the question of how the vdW-DF
performs for solids and covalently bonded systems such as a CO$_2$
molecule or crystalline Si. In Sec.~\ref{sec:nature} we use the vdW-DF
self-consistent charge density to discuss the nature of the van der
Waals bond. We conclude in Sec.~\ref{sec:conclusions}. Some details of
the derivation of the potential are collected in
Appendices~\ref{sec:details_a}, \ref{gradient}, and 
\ref{sec:details_b}.

%%%%%%%%%%%%%%%%%%%%%%%%%%%%%%%%%%%%%%%%%%%%%%%%%%%%%%%%%%%%%%%%%%%%%%%%
\section{Derivation of the exchange-correlation potential}
\label{sec:derivative}
%%%%%%%%%%%%%%%%%%%%%%%%%%%%%%%%%%%%%%%%%%%%%%%%%%%%%%%%%%%%%%%%%%%%%%%%

The density functional in question is defined in
Ref.~[\onlinecite{dion04}]. It consists of several different
contributions, i.e.
\begin{equation}\label{equ:functional}
E_{xc}[n] = E^{\text{revPBE}}_x[n] + E^{\text{LDA}}_c[n] + E^{\text{nl}}_c[n]\;.
\end{equation}
The first two parts are simply revPBE exchange and LDA correlation. We
note that these parts represent LDA plus gradient corrections and are
hence \emph{semilocal}. The van der Waals interaction enters through a
fully \emph{nonlocal} correction $E^{\text{nl}}_c[n]$ that concerns the
correlation only. In the following we are interested in the
exchange-correlation potential that originates from this nonlocal
contribution. However, we have to keep in mind that the total
exchange-correlation potential corresponding to
Eq.~(\ref{equ:functional}) also includes the parts stemming from
$E^{\text{revPBE}}_x[n]$ and $E^{\text{LDA}}_c[n]$---these expressions
are well-known and can be found elsewhere.\cite{PBEvx,LDAvc}

In order to calculate the potential of interest, we have to take the
functional derivative of the energy with respect to the density, i.e.
\begin{eqnarray}\label{twopoint}
E_c^\text{nl}[n]     &=& \frac{1}{2}\int\dr d^3r'\,
                         n(\rr)\phi(\rr,\rr')n(\rr\,')\;,\label{twopointE}\\
v^{\text{nl}}_c(\rt) &=& \frac{\delta E^{\text{nl}}_c[n]}
                         {\delta n(\rt)}\;,\label{twopointv}
\end{eqnarray}
where $\phi(\rr,\rr')$ is a function depending on $\rr -\rr'$ and the
electronic densities $n$ and the magnitude of their gradients at the
points  $\rr$ and $\rr'$. Details about the kernel $\phi(\rr,\rr')$ and
its evaluation can be found in Appendix~\ref{sec:details_a}. It follows
that
\begin{eqnarray}\label{equ:v_xc_10}
v^{\text{nl}}_c(\rt) &=&\;\;\;
   \frac{1}{2}\int\dr\dr' \: \frac{\delta n(\rr)}{\delta n(\rt)}
   \phi(\rr,\rr')n(\rr')\nonumber\\
&& +\frac{1}{2}\int\dr\dr' \: n(\rr)\phi(\rr,\rr')
   \frac{\delta n(\rr')}{\delta n(\rt)}\\
&& +\frac{1}{2}\int\dr\dr' \: n(\rr)\frac{\delta\phi(\rr,\rr')}
   {\delta n(\rt)}n(\rr')\nonumber\;.
\end{eqnarray}
The first two lines can be simplified if we use
\begin{equation}
\frac{\delta n(\rr)}{\delta n(\rt)} = \delta(\rr-\rt)\;.
\end{equation}
It turns out that the kernel $\phi(\rr,\rr')$ depends on $\rr$ and
$\rr'$ only through two functions $d$ and $d'$
\begin{subequations}\label{equ:d_def}
\begin{eqnarray}
d(\rr,\rr')  &=& |\rr-\rr'|\,q_0(\rr) = R_{\rr\rr'}q_0(\rr)\\
d'(\rr,\rr') &=& |\rr-\rr'|\,q_0(\rr')= R_{\rr\rr'}q_0(\rr')\;,
\end{eqnarray}
\end{subequations}
where we have introduced $\RR_{\rr\rr'}=\rr-\rr'$ and
$R_{\rr\rr'}=|\RR_{\rr\rr'}|$.  The definition of $q_0(\rr)$ can be
found in Eqs.~(11) and (12) of Ref.~[\onlinecite{dion04}] which reads
\begin{equation}
q_0(\rr) = -\frac{4\pi}{3}\varepsilon_{\text{xc}}^{\text{LDA}}
    n(\rr)-\frac{Z_{ab}}{9}s^2(\rr)\kF(\rr)\;,
\label{q0}
\end{equation}
where we have used the standard expressions for the Fermi wave vector
and the reduced gradient 
\begin{equation}
k_{\text{F}}^3(\rr) = 3\pi^2n(\rr)\;,\qquad
\redgrad(\rr) = \displaystyle\frac{\nabla n(\rr)}{2\kF(\rr)n(\rr)}\;,
\end{equation}
and $Z_{ab}=-0.8491$. A complete discussion of the gradient term
in \eq{q0} is given in Appendix \ref{gradient}.

We may now write 
\begin{equation}
\phi(\rr,\rr') = \phi\big(d(\rr,\rr'),d'(\rr,\rr')\big) =
\phi(d,d') = \phi(d',d)\;.
\end{equation}
The symmetry in the last equality is apparent by exploiting the
definition of $\phi(\rr,\rr')$ in Appendix~\ref{sec:details_a}. In order
to calculate the derivative in the third line of Eq.~(\ref{equ:v_xc_10})
we can use this symmetry and find
\begin{equation}
\frac{\delta\phi(\rr,\rr')}{\delta n(\rt)} = 
\frac{\partial \phi(d,d')}{\partial d}\frac{\delta d}{\delta n(\rt)}
+\frac{\partial\overbrace{\phi(d,d')}^{\phi(d',d)}}
{\partial d'}\frac{\delta d'}{\delta n(\rt)}\;.
\end{equation}
If we keep in mind that both terms appear under a double integral, we
can change the integration variables and find that these two terms are
equivalent. Then, simply inserting a factor of two and abbreviating
$\frac{\partial \phi(d,d')}{\partial d} = \phi_d(d,d') =
\phi_d(\rr,\rr')$ yields
\begin{eqnarray}\label{equ:v_xc_20}
v^{\text{nl}}_c(\rt) &=& \;\;\;\int\dr' \: \phi(\rt,\rr')n(\rr')\\
&& + \int\dr\dr' \: n(\rr)\phi_d(\rr,\rr')
   \frac{\delta d(\rr,\rr')}{\delta n(\rt)}n(\rr')\;.\nonumber
\end{eqnarray}

The missing derivative in the equation above follows from the definition
of $d(\rr,\rr')$ in Eq.~(\ref{equ:d_def}), i.e. 
\begin{equation}\label{equ:dd_dn}
\frac{\delta d(\rr,\rr')}{\delta n(\rt)} =
R_{\rr\rr'}\frac{\delta q_0(\rr)}{\delta n(\rt)}\;.
\end{equation}
If we denote the derivative of $\varepsilon_{\text{xc}}^{\text{LDA}}$
with respect to the density as $\varepsilon_{\text{xc}}^{\text{LDA}'}$
and simplify further, it follows
\begin{eqnarray}\label{equ:dq_dn}
\frac{\delta q_0(\rr)}{\delta n(\rt)} &=& {}-\frac{4\pi}{3}
   \varepsilon_{\text{xc}}^{\text{LDA}'}(\rr) \:
   \delta(\rr-\rt)\\
&& {}-\frac{Z_{ab}}{9}\frac{1}{n(\rr)} \: \redgrad(\rr)\cdot
   \nabla\delta(\rr-\rt)\nonumber\\
&& {}+\frac{7}{3}\frac{Z_{ab}}{9}\frac{1}{n(\rr)} \: s^2(\rr)\kF(\rr)
   \: \delta(\rr-\rt)\;.\nonumber
\end{eqnarray}
Equation~(\ref{equ:dq_dn}) together with Eq.~(\ref{equ:dd_dn}) can now
be inserted in the expression for the potential,
Eq.~(\ref{equ:v_xc_20}), and it follows
\begin{eqnarray}\label{eqi:v_xc_30}
\lefteqn{v^{\text{nl}}_c(\rt) = \int\dr' \: \phi(\rt,\rr')n(\rr')}\\
&& {} - \frac{4\pi}{3}\int\dr' \: n(\rt)\phi_d(\rt,\rr')R_{\rt\rr'}
 \varepsilon_{\text{xc}}^{\text{LDA}'}(\rt)n(\rr')\nonumber\\
&& {} - \frac{Z_{ab}}{9}\int\dr\dr' \: \phi_d(\rr,\rr')R_{\rr\rr'}
   \redgrad(\rr)\cdot\big(\nabla\delta(\rr-\rt)\big)n(\rr')\nonumber\\
&& {} + \frac{7}{3}\frac{Z_{ab}}{9}\int\dr' \: \phi_d(\rt,\rr')R_{\rt\rr'}
   s^2(\rt)\kF(\rt)n(\rr')\;.\nonumber
\end{eqnarray}
The third line in this equation includes the gradient of the
$\delta$-function. Partial integration lets us rewrite the
corresponding integral as
\begin{equation*}
+ \frac{Z_{ab}}{9}\int\dr' \: \nabla_\rt\cdot\Big[\phi_d(\rt,\rr')R_{\rt\rr'}
\redgrad(\rt)\Big]n(\rr')\;,
\end{equation*}
where the sufficiently rapid fall-off of $\phi_d$ (see
Eq.~\ref{equ:phiasymp}) guaranties the vanishing of the integrated part.
All contributions now have the common form $\int\dr' \dots \, n(\rr')$
so that, after dropping the tilde $(\rt\to\rr)$, we may simply
write:
\begin{eqnarray}\label{equ:v_xc_40}
\lefteqn{v^{\text{nl}}_c(\rr) = \int\dr' \: n(\rr')\;\times}\\
&& \bigg[\phi(\rr,\rr') + 
   \frac{Z_{ab}}{9}\nabla\cdot\Big[\phi_d(\rr,\rr')R_{\rr\rr'}
   \redgrad(\rr)\Big]\nonumber\\
&& {} + \phi_d(\rr,\rr')R_{\rr\rr'}\Big(\frac{7}{3}
   \frac{Z_{ab}}{9}s^2(\rr)\kF(\rr) - \frac{4\pi}{3}n(\rr)
   \varepsilon_{\text{xc}}^{\text{LDA}'}(\rr)\Big)\bigg]\;.\nonumber
\end{eqnarray}

The term including the gradient $\nabla\cdot[\dots]$ in the equation
above can be further simplified, but the details are deferred to 
Appendix~\ref{sec:details_b}. If we use Eq.~(\ref{equ:d_def}) to
replace $R_{\rr\rr'}$ in favor of $d(\rr,\rr')$ or $d'(\rr,\rr')$ and
collect similar terms, we find as the final result
\begin{equation}\label{equ:vfinal}
v^{\text{nl}}_c(\rr) = \int\dr' \: n(\rr')\sum_{i=1}^4
\alpha_i(\rr,\rr') \: \Phi_i(\rr,\rr')\;,
\end{equation}
where the functions $\alpha_i(\rr,\rr')$ and $\Phi_i(\rr,\rr')$ are
given by:
\begin{subequations}\label{equ:alpha}
\begin{eqnarray}
\alpha_1 &=&\frac{1}{q_0(\rr)}\Big[
   \frac{Z_{ab}}{9}\nabla\cdot\redgrad(\rr)+
   \frac{7}{3}\frac{Z_{ab}}{9}s^2(\rr)\kF(\rr)
   \nonumber\label{equ:alpha1}\\
&& \qquad\quad -\frac{4\pi}{3}n(\rr)
   \varepsilon_{\text{xc}}^{\text{LDA}'}(\rr)\Big]\\
\alpha_2 &=& \frac{Z_{ab}}{9} \frac{\redgrad(\rr)\cdot\nabla q_0(\rr)}
   {q_0(\rr)^2}\label{equ:alpha2}\\
\alpha_3 &=& \frac{Z_{ab}}{9}\hat{\RR}_{\rr\rr'}\cdot\redgrad(\rr)
   \label{equ:alpha3}\\
\alpha_4 &=& 1\label{equ:alpha4}
\end{eqnarray}
\end{subequations}
and
\begin{subequations}\label{equ:phi}
\begin{eqnarray}
\Phi_1 &=& d\phi_d(d,d')\label{equ:phi1}\\
\Phi_2 &=& d^2\phi_{dd}(d,d')\label{equ:phi2}\\
\Phi_3 &=& \phi_d(d,d') + d\phi_{dd}(d,d')+d'\phi_{dd'}(d,d')\label{equ:phi3}\\
\Phi_4 &=& \phi(d,d')\label{equ:phi4}\;,
\end{eqnarray}
\end{subequations}
where $\Phi_4$ is the kernel function defined in
Ref.~[\onlinecite{dion04}].
Here, we have abbreviated higher order derivatives of $\phi(d,d')$ as
previously,
\begin{equation*}
\frac{\partial^2\phi(d,d')}{\partial d\partial d}  = \phi_{dd}(d,d')
\quad\text{and}\quad
\frac{\partial^2\phi(d,d')}{\partial d\partial d'} = \phi_{dd'}(d,d')\;.
\end{equation*}
For completeness, details about the evaluation of $\phi_d(d,d')$,
$\phi_{dd}(d,d')$, and $\phi_{dd'}(d,d')$ are provided in
Appendix~\ref{sec:details_a}.

\begin{figure}
\includegraphics[width=\columnwidth]{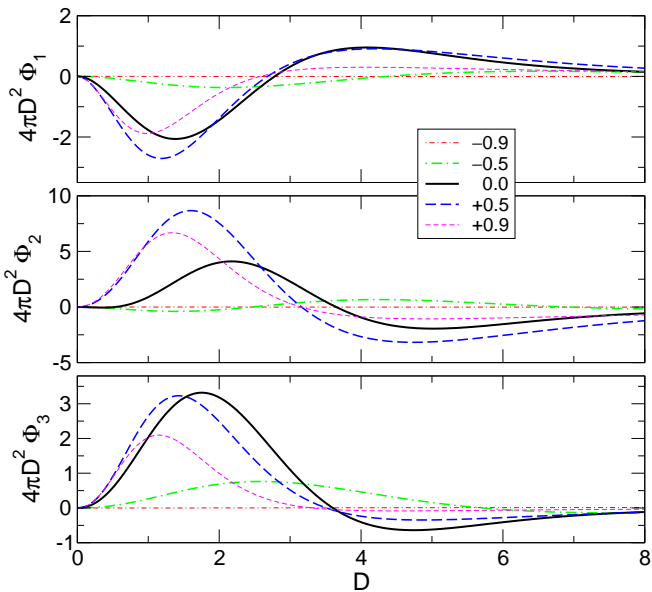}
\caption{\label{fig:kernels} The kernels $\Phi_1$, $\Phi_2$, and 
$\Phi_3$ given in \eqs{equ:phi} are plotted as a function of $D$ for
values of $\delta=\pm0.9,\pm0.5$, and 0.0 [see \eqs{equ:sumdiff}].}
\end{figure}

The three extra internal kernel functions $\Phi_1(d,d')$ through
$\Phi_3(d,d')$ in Eqs.~(\ref{equ:phi1}--\ref{equ:phi3}) necessary for
$v_c^{\text{nl}}$ are the analogues of the single function $\phi(d,d')$
($\equiv\Phi_4(d,d')$) used for $E_c^\text{nl}$. However, unlike the
latter, they are not symmetric under the exchange of $d$ and $d'$.
Consequently, when expressed in terms of the sum and difference
variables $D$ and $\delta$ defined by\cite{dion04}
\begin{subequations}
\label{equ:sumdiff}
\begin{eqnarray} 
d &=&D(1+\delta)\;, \label{equ:sum}\\
d'&=&D(1-\delta)\;, \label{equ:diff}
\end{eqnarray}
\end{subequations}
their values depend on the sign of $\delta$. Plots of these three new
kernels are given in Fig.~\ref{fig:kernels}.  Comparison with Fig.~1 of
Ref.~[\onlinecite{dion04}] (see erratum), shows that there are no
unexpected features. The details of the evaluation are given in Appendix
\ref{sec:details_a}.
The asymptotic form of the kernels for $v^{\text{nl}}_c$  for large
$R\equiv|\rr-\rr'|$ may be determined by using the 
relation\cite{dion04} 
\begin{equation}
\phi(d,d')\rightarrow-\frac{12(4\pi/9)^3}{d^2d'^2(d^2+d'^2)}\;.
\label{equ:phiasymp}
\end{equation}
The use of Eq.~(\ref{equ:phiasymp}) in  Eqs.~(\ref{equ:phi}) plus the
fact that $D\propto R^1$ and $\delta\propto R^0$ gives the asymptotic
dependence $R^{-6}$ to $\Phi_1$, $\Phi_2$, and $\Phi_4$, while $\Phi_3$
falls off faster and thus does not contribute to the asymptote.

The asymptotic form above follows from the form of the model response
function that determines the kernels. It gives the correct asymptotic
dependence for all finite sized fragments, and for all infinite
fragments that are not metallic. A couple of decades ago,
Barash\cite{barash} showed that infinite metallic fragments of reduced
dimensionality could show a more slowly decaying asymptote than would be
given by the appropriate integrals over (\ref{equ:phiasymp}). This
effect occurs because the dipole response functions of infinite metals
may not be sufficiently convergent at small frequency and wave vector. 
A number of examples of special cases have been worked out
more recently.\cite{metal} We can say with certainty that, with the response
function presently used, our functional will fail for infinite  metallic
(or semimetallic) fragments of reduced dimensionality at very large
distances (for example, parallel metallic sheets). The distance where
the crossover occurs will depend on the details of the material. We are
unaware of any such material where this distance is known.

The arrangement of the results for $v^{\text{nl}}_c$ in \eq{equ:vfinal}
according to \eqs{equ:alpha} and (\ref{equ:phi}) was chosen to
facilitate rapid numerical evaluation. For a grid of $N$ points, the
evaluation of $v^{\text{nl}}_c$ on each is $\mathcal{O}(N^2)$ according
to \eq{equ:vfinal}. However, the evaluation of the  $\Phi$'s in
\eqs{equ:phi} needs only to be done once to create an interpolation
table. The conversion between $\rr$ ($\rr'$) and $d$ ($d'$)
[\eqs{equ:d_def}] of course needs to be done $\mathcal{O}(N^2)$ times,
but the lengthy part of that would be the calculation of $q_0$, which
instead can be calculated in advance on each grid point, i.e. it is
$\mathcal{O}(N)$. The calculation of $\alpha_1$ and $\alpha_2$
[\eqs{equ:alpha}] is $\mathcal{O}(N)$.  So, the only parts of the
calculation that are $\mathcal{O}(N^2)$ are the calculation of the
components of $\RR$ for $\alpha_3$ [\eq{equ:alpha3}], the calculation
of  $|\RR|$  for $\rr$ to $d$ conversion [\eqs{equ:d_def}], two table
interpolations, and a few arithmetic operations.

%%%%%%%%%%%%%%%%%%%%%%%%%%%%%%%%%%%%%%%%%%%%%%%%%%%%%%%%%%%%%%%%%%%%%%%%
\section{Results}
\label{sec:results}
%%%%%%%%%%%%%%%%%%%%%%%%%%%%%%%%%%%%%%%%%%%%%%%%%%%%%%%%%%%%%%%%%%%%%%%%

The exchange-correlation potential derived in the previous section will
now be applied to various systems. Calculations labeled as
self-consistent (SC) include the potential (\ref{twopointv}) as part of
the Kohn-Sham potential, so that the density used to evaluate the
nonlocal correlation energy, Eq.~(\ref{twopointE}), is fully
self-consistent. Results in which the nonlocal vdW correlation energy
(\ref{twopointE}) is calculated  using the density from a
self-consistent PBE calculation will be referred to as non-SC. It should
be noted that all previous vdW-DF calculations\cite{dion04, rydberg03,
langrethIJQC, aaron05, Thonhauser06, svetla2, svetla1,
tsuzuki-naphthalene, benz-graph, pe} were preformed with such a non-SC
post-process procedure.

It is apparent that the post-process evaluation of the vdW-DF is an
approximation since the charge density is not allowed to evolve under
the full vdW-DF functional.  Although there are many arguments that the
effect of this approximation is small, the ultimate test is obviously a
comparison of our new SC results with the older non-SC results. In the
following we shall make this comparison.

In addition to the SC and non-SC results, we will also show results
marked as ``no $E_c^{\text{nl}}$''. These results were obtained by only
considering $E^{\text{revPBE}}_x + E^{\text{LDA}}_c$ in
Eq.~(\ref{equ:functional}) and neglecting the nonlocal part
$E_c^{\text{nl}}$. This will allow us to prove that the correct behavior
of the vdW-DF for van der Waals systems indeed is encoded in the
nonlocal correlation $E_c^{\text{nl}}$. The calculations below were made
with the aid of the plane-wave code ABINIT\cite{ABINIT} and the
real-space code PARSEC.\cite{PARSEC}

%%%%%%%%%%%%%%%%%%%%%%%%%%%%%%%%%%%%%%%%%%%%%%%%%%%%%%%%%%%%%%%%%%%%%%%%
\subsection{Ar and Kr dimers}

\begin{figure}
\includegraphics[width=\columnwidth]{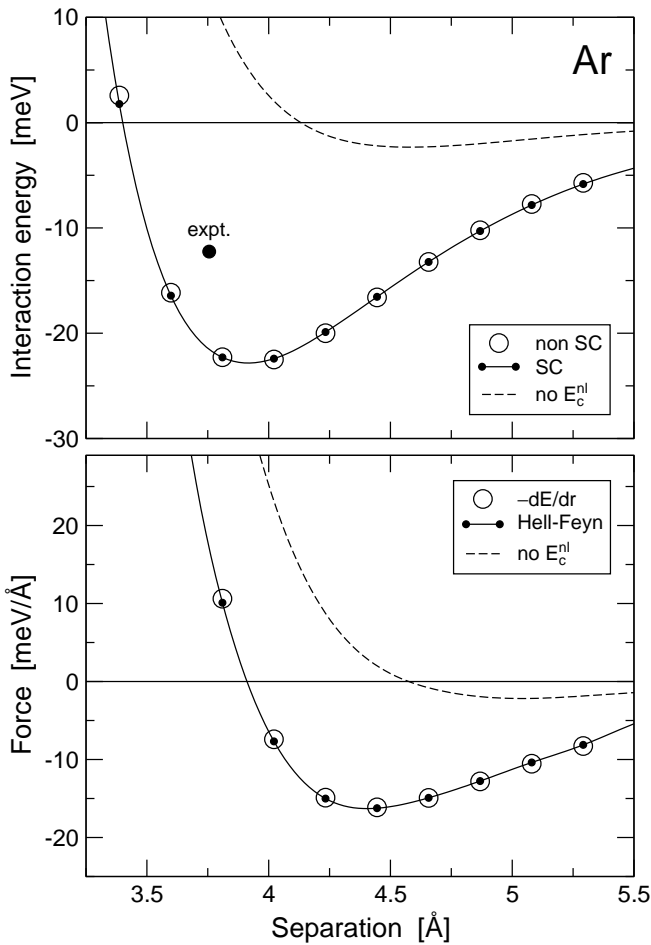}
\caption{\label{fig:Ar}{\bf(top)} Interaction energy of the Ar dimer as
a function of separation. Plotted are self-consistent and non
self-consistent results. In addition, we show results where
$E^{\text{nl}}_c$ has been neglected. Experimental values are taken from
Ref.~[\onlinecite{ogilvie92}]. {\bf(bottom)} Forces calculated as the
derivative of the energy ($-dE/dr$) and the Hellmann-Feynman forces.}
\end{figure}

\begin{figure}
\includegraphics[width=\columnwidth]{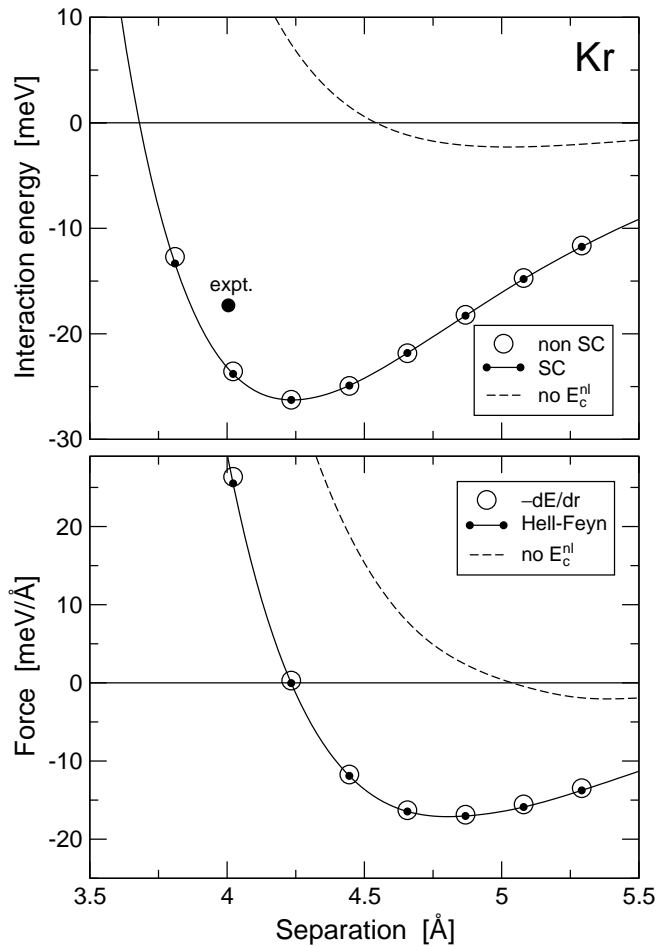}
\caption{\label{fig:Kr} Same as Fig.~\ref{fig:Ar} except here describing
the Kr dimer.}
\end{figure}

We start out by applying the self-consistent vdW-DF to some rare gas
dimers, which, due to their closed valence shells, owe much of their
binding to van der Waals interactions. Results for the interaction
energy of an Ar and a Kr dimer as a function of separation are depicted in
Figs.~\ref{fig:Ar} and \ref{fig:Kr}.
It can be seen that the differences between SC and non-SC calculations
are indeed very small, and on the scale of this plot they are
negligible. In general, these differences are larger for smaller
separations, where the perturbation from the nonlocal term is larger. 
For Ar and Kr we find the binding distances to be 3.9~\AA\ and 4.2~\AA,
which is approximately 5\% larger than their corresponding experimental
values.\cite{ogilvie92} On the other hand, our interaction energies are
stronger than experiment. This behavior of the binding distance is
consistent with all previous
calculations\cite{rydberg03,dion04,langrethIJQC,aaron05,Thonhauser06,
svetla2,svetla1,tsuzuki-naphthalene,benz-graph,pe} and  has been 
attributed to the form of the exchange used.\cite{aaron05, Thonhauser06}
In addition, it can be seen that without the nonlocal contribution (no
$E_c^{\text{nl}}$) the binding is much too weak and the separation is
too large. In some cases, as we shall see later in the text, without
this contribution binding does not occur at all. It thus becomes obvious
that the addition of the nonlocal contribution $E^{\text{nl}}_c$
dramatically improves the description of these van der Waals systems.

As a result of our DFT calculations, we not only obtain the energies
discussed above, but also the corresponding Hellmann-Feynman forces
which act on each nuclei. These are simply the electrostatic forces,
which require the correct self-consistent charge distribution for their
determination. Until now, it was impossible to obtain them within
vdW-DF. Results for these forces are plotted in the bottom panels of
Figs.~\ref{fig:Ar} and \ref{fig:Kr}. In order to verify the correctness
of these forces, we have plotted the negative derivative of the energy
from above ($-dE/dr$). It can be seen that the zero point of the forces
perfectly aligns with the corresponding energy minimum.  As for the
energy, we have again included results for ``no $E_c^{\text{nl}}$''. It is now
apparent that semilocal exchange-correlation alone cannot
describe van der Waals systems correctly and most of the binding force
originates in the nonlocal part of the functional.

%%%%%%%%%%%%%%%%%%%%%%%%%%%%%%%%%%%%%%%%%%%%%%%%%%%%%%%%%%%%%%%%%%%%%%%%
\subsection{CO$_2$ dimer}
\label{sec:CO2_dimer}

Next, we apply the vdW-DF to a slightly more complex molecule, i.e. the
CO$_2$ dimer. However, before we discuss the details about the dimer, it
is interesting to first explore the behavior of the vdW-DF for a single
CO$_2$ molecule. The CO$_2$ molecule is a covalently bonded system, with
only minor contributions from vdW interactions. As such, it is
interesting to investigate how vdW-DF treats such a system. The
important question here is whether or not the functional will continue
to give good results for cases in which van der Waals interactions are
negligible or non-existent.

For CO$_2$ we can partly answer this question by calculating the 
optimal C--O bond length. Our results show that the bond length that
vdW-DF predicts is almost identical to the bond length obtained from a
standard PBE calculation. It turns out that the nonlocal part of the
functional ($E_c^{\text{nl}}$) only makes a negligible
contribution to the total energy. In this regard, the total
exchange-correlation energy computed with the vdW-DF is just the sum of
the $E^{\text{revPBE}}_x[n]$ and $E^{\text{LDA}}_c[n]$, i.e. revPBE
exchange and LDA correlation. It remains to be determined whether this
form of exchange-correlation energy is appropriate for the system of
interest, but the main point is that in the regime of weak van der Waals
forces, the vdW-DF will not change or alter well established DFT
results. We will readdress the question of covalently bonded systems and
vdW-DF in the context of a solid in Sec.~\ref{sec:silicon}.

\begin{figure}
\includegraphics[width=\columnwidth]{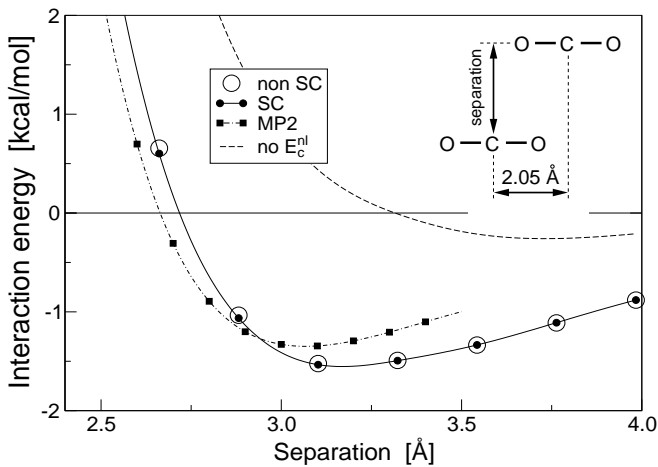}
\caption{\label{fig:CO2}Interaction energy of a CO$_2$ dimer as a
function of its separation at a fixed slip distance of 2.05~\AA. The
inset shows the CO$_2$ dimer geometry.}
\end{figure}

Now we turn to the bonding of the CO$_2$ dimer. We have performed
calculations for the interaction energy of this dimer as a function of
the dimer separation and slip distance. We find that the preferred
parallel slip distance is 2.05~\AA, and the corresponding interaction
energy as a function of dimer separation is depicted in
Fig.~\ref{fig:CO2}. Similar to Ar and Kr, the results indicate that the
differences between non-SC and SC vdW-DF calculations are negligible.
Furthermore, the C--C separation distance of 3.772~\AA\ is in good
agreement with the experimentally determined separation distance of
3.602~\AA.\cite{Walsh87p265}  The vdW-DF interaction energy of 1.55
kcal/mol is comparable to the MP2 value of 1.36 kcal/mol (extrapolated
to the complete basis set limit)\cite{Tsuzuki98p2169} as well as the
value of 1.60 kcal/mol determined by Becke and coworkers using their
density-functional model for the dispersion
interaction.\cite{Becke05p154101} In addition, Fig.~\ref{fig:CO2} shows
that the equilibrium separation distance for the CO$_2$ dimer is
slightly larger than the MP2 results.\cite{Tsuzuki98p2169}

As before, by looking at the ``no $E_c^{\text{nl}}$'' curve we can see that
most of the binding originates in the nonlocal part of the
functional.

%%%%%%%%%%%%%%%%%%%%%%%%%%%%%%%%%%%%%%%%%%%%%%%%%%%%%%%%%%%%%%%%%%%%%%%%
\subsection{Water on benzene}

\begin{figure}
\includegraphics[width=\columnwidth]{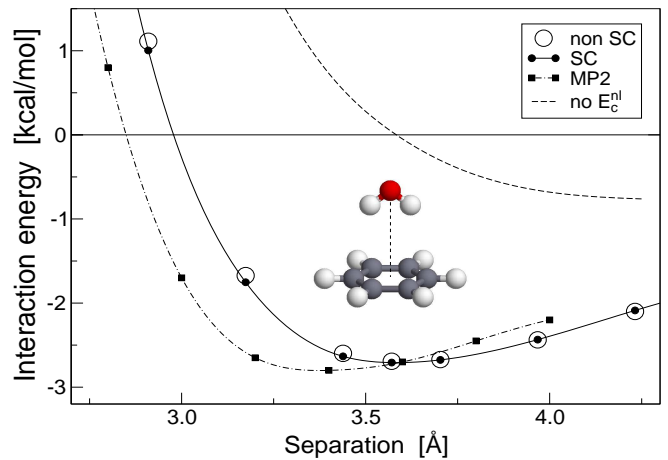}
\caption{\label{fig:benzene}Interaction energy of water on benzene.
The little inset shows the geometry.}
\end{figure}

We now move to the physically and chemically more interesting question
of how a typical solvent like water binds to hydrocarbons.  Here, we
consider the simple test case of the interaction energy between water
and benzene. We limit ourselves to a particular
configuration, in which the water molecule is positioned with the
hydrogen atoms at equivalent heights pointing towards the benzene
plane.  The oxygen atom is positioned directly above the center of the
benzene molecule, as depicted in the little inset in
Fig.~\ref{fig:benzene}.

First, we relaxed the bond length of the water molecule and benzene ring
individually. The optimal H--O bond length for the water molecule and
the H--O--H angle were found to be 0.973 \AA\ and 106$^{\circ}$,
respectively, in quite good agreement with the experimental values of
0.958 \AA\ and 104.5$^{\circ}$.\cite{webbook} The C--C and C--H bond
lengths for the benzene molecule were 1.396 \AA\ and 1.100 \AA, where
experiments find 1.397 \AA\ and 1.084 \AA, respectively.\cite{webbook}
Our results for the interaction energy versus the vertical separation
distance between the oxygen atom and the center of the benzene ring is
plotted in Fig.~\ref{fig:benzene}. Again, we see that the difference
between SC and non-SC calculations is very small. Consistent with our
previous results, the difference is more evident at shorter separation
distances. For comparison, we plotted the results from a recent MP2
study.\cite{Zimmerli04} We find fair agreement, with MP2
calculations giving a shorter equilibrium distance and lower interaction
energies than vdW-DF. The difference for both is less than 5\%.

Once again, we find that without $E_c^{\text{nl}}$, the complex binds at
a much too large distance with much weaker binding energies---supporting the
fact that the nonlocal correlation energy is crucial for the correct
description of van der Waals binding.

%%%%%%%%%%%%%%%%%%%%%%%%%%%%%%%%%%%%%%%%%%%%%%%%%%%%%%%%%%%%%%%%%%%%%%%%
\subsection{Cytosine dimer}

The applicability of vdW-DF to monosubstituted benzene dimers has been
shown in Ref.~[\onlinecite{Thonhauser06}]. We will now extend this work
by considering cytosine, a simple nucleic acid base.

\begin{figure}
\includegraphics[width=\columnwidth]{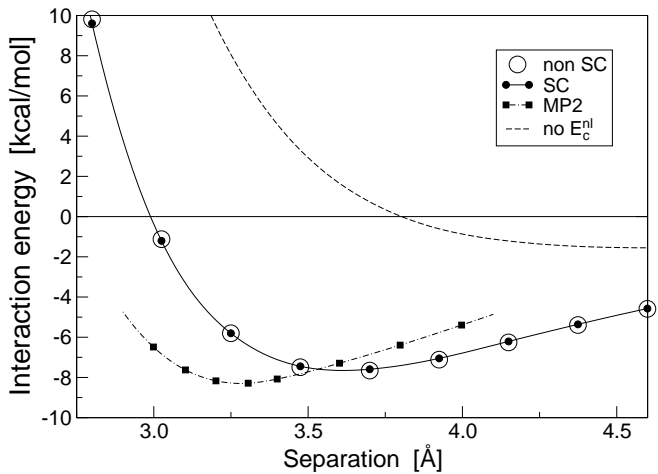}
\caption{\label{fig:cytosine}Interaction energy of a cytosine
dimer as a function of separation.}
\end{figure}

We study the interaction energy of a cytosine dimer as a function of
separation (Fig.~\ref{fig:cytosine}). The geometry is such that the two
molecules are placed on top of each other with one of them
rotated by 180$^\circ$ around an axis that passes through the center of
both rings. Consistent with other findings reported in this paper, there
are no significant differences between the SC and non-SC results.
Similarly as in the cases before, the ``no $E_c^{\text{nl}}$'' curve
reveals that the majority of the binding energy comes from the nonlocal
contribution of the vdW-DF. As usual, while the comparison to MP2
calculations is quite good, we find a slightly larger binding distance.

%%%%%%%%%%%%%%%%%%%%%%%%%%%%%%%%%%%%%%%%%%%%%%%%%%%%%%%%%%%%%%%%%%%%%%%%
\subsection{Crystalline Silicon}
\label{sec:silicon}

We now choose to apply the new functional to an extended system and return
to the question raised in Sec.~\ref{sec:CO2_dimer}. We
purposefully selected an extended system in which the bonding is mostly
covalent and where van der Waals interactions should be unimportant.
Again, we seek to understand how the vdW-DF treats such systems.

\begin{figure}
\includegraphics[width=\columnwidth]{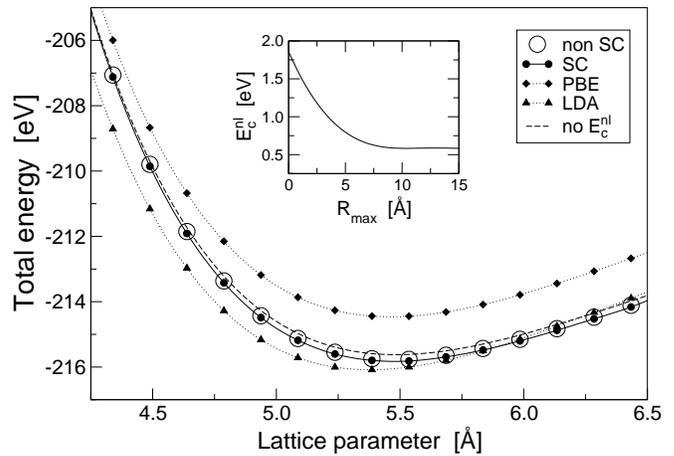}
\caption{\label{fig:Si}Total energy (per two-atom unit cell) of Si as a
function of the lattice constant. The inset shows the convergence of
$E_c^{\text{nl}}$ with respect to the cutoff radius $R_{\text{max}}$.
See text for further details.}
\end{figure}

In Fig.~\ref{fig:Si}, we plot the total energy (per two-atom unit cell)
of crystalline silicon as a function of the lattice parameter. In
addition to the ``usual'' curves for SC and non-SC results, which show
no differences, we now include results from calculations performed with
standard LDA and PBE functionals. The main point of Fig.~\ref{fig:Si} is
that all calculations result in very similar lattice constants: non-SC
(5.48 \AA), SC (5.48 \AA), no $E_c^{\text{nl}}$ (5.48 \AA), PBE (5.46
\AA), and LDA (5.37 \AA).  In other words, the nonlocal contribution in
systems like silicon is indeed negligible and the functional performs as
expected for the local and semilocal contributions, i.e.\ revPBE
exchange and LDA correlation. This is even more evident in the fact that
the lines for SC and ``no $E_c^{\text{nl}}$'' almost coincide, resulting
in exactly the same lattice constant.

At this point, we would also like to mention another important aspect of
periodic systems. Since crystalline Si is an extended system,
interactions between different unit cells have to be included when the
energy in Eq.~(\ref{equ:functional}) and the potential in
Eq.~(\ref{equ:vfinal}) are evaluated. From Fig.~\ref{fig:kernels} it is
apparent that all kernel functions decrease in absolute magnitude with
increasing $D$, and in turn, also with increasing separation
$|\rr-\rr'|$ of the two points $\rr$ and $\rr'$. Thus, it is natural to
introduce a maximum distance $R_\text{max}$ above which the contribution
of a given pair $\rr$ and $\rr'$ is no longer included. For the case of
Si, the convergence of $E_c^{\text{nl}}$ with respect to this cutoff
radius $R_{\text{max}}$ is depicted in the inset in Fig.~\ref{fig:Si}. 
It can be seen that the convergence is quicker than for materials
containing carbon.\cite{benz-graph,pe} For the purpose of this study we
have chosen $R_{\text{max}}=15$~\AA .

The findings for crystalline silicon parallel the earlier findings for
the CO$_2$ molecule in Sec.~\ref{sec:CO2_dimer}: When the van der
Waals interactions in an extended system become negligible, the
contribution of the nonlocal part $E_c^{\text{nl}}$ to the total energy
becomes negligible. Again, vdW-DF does not affect well established DFT
results. With this, we conclude our quantitative results and move on to
a qualitative description of the van der Waals bond.

%%%%%%%%%%%%%%%%%%%%%%%%%%%%%%%%%%%%%%%%%%%%%%%%%%%%%%%%%%%%%%%%%%%%%%%%
\section{Nature of the van der Waals bond}
\label{sec:nature}
%%%%%%%%%%%%%%%%%%%%%%%%%%%%%%%%%%%%%%%%%%%%%%%%%%%%%%%%%%%%%%%%%%%%%%%%

The conventional picture of the van der Waals interaction envisions in
its simplest form two fragments with distinct charge distributions,
whose motions are correlated so as to reduce the interfragment
electron-electron repulsion and hence produce a net attractive force
between the fragments. For distant fragments the dipolar interactions
caused by these correlations are dominant, and one obtains the familiar
attractive $r^{-6}$ interaction (at \textit{very} large distances,
typically 100 {\AA} or more, this interaction is further weakened by
relativistic retardation effects). The vdW-DF method embraces this
picture, which indeed was central to its derivation, and also includes
the contribution of more complex correlated motions that occur as the
fragments get closer and even merge.

However, the van der Waals bond must also have a completely different
feature as part of its nature.  This feature arises because the nuclei
are classical particles influenced only by Coulomb forces and immune to
the fluctuations in such forces due to the more rapid electronic
motions. Because of the stationary property of the energy with respect
to variations in the wave function, these forces are easily
calculated---a result often attributed to Hellmann\cite{hellmann} and
Feynman,\cite{feynman} but whose foundation is much
older.\cite{ehrenfest,pauli} The consequence is that the static or
ground state electronic charge distribution must deform itself in such a
way as to produce the required forces on the nuclei by classical Coulomb
interactions alone. This concept is familiar for covalent bonds, but how
is it implemented for van der Waals bonds?

\begin{figure}
\includegraphics[width=0.9\columnwidth]{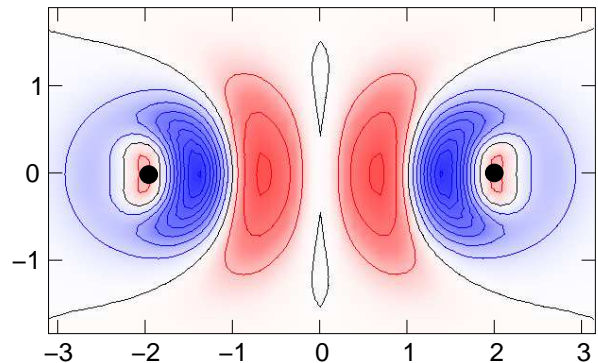}
\caption{\label{fig:dens}The bonding charge of a van der Waals complex.
Shown is the difference in induced electron density
$n^{\text{full}}_{\text{ind}}-n_{\text{ind}}^{\text{no nl}}$ for the Ar
dimer with a separation of 4.02 \AA. The scale is in \AA\ and the black
dots mark the position of the nuclei. The zero level is marked by the
black contour. Red areas represent areas of electron density gain when
the nonlocal part is included; conversely, blue areas indicate loss of
electron density. Increments between contour lines are
5$\times$10$^{-5}$ electrons/\AA$^3$.}
\end{figure}

We can now answer this question by calculating how the static charge
density changes when the nonlocal contribution $E_c^{\text{nl}}$ is
included in the functional. To this end, we investigate the density of
the Ar dimer in a plane that includes both atoms. We calculate the
induced electron density that occurs when the atoms bond, i.e.
$n_{\text{ind}}=n_{\text{bond}}-n_{\text{atom}}$, where
$n_{\text{bond}}$ is the density of the dimer and $n_{\text{atom}}$ is
the density of the isolated atoms. We can now study the difference in
the induced electron density when the full functional is used
$n_{\text{ind}}^{\text{full}}$, compared to the induced density that
results without the nonlocal part $n_{\text{ind}}^{\text{no nl}}$.
Results for $n^{\text{full}}_{\text{ind}}-n_{\text{ind}}^{\text{no nl}}$
for a separation of 4.02 \AA\ are depicted in Fig.~\ref{fig:dens}.
The plot shows the change in the induced electron density when the
nonlocal part of the functional is ``turned on.'' As expected, it can be
seen that the electron density from around the nuclei moves between the
atoms, which in turn explains the stronger binding in terms of implied
changes in the electrostatic forces on the nuclei arising from this
charge redistribution.  Although the change in density is very small
($\sim 10^{-4}$ electrons/\AA$^3$), the resulting binding energies and
forces change considerably, as seen in Figs.~\ref{fig:Ar} and
\ref{fig:Kr}.  This is the van der Waals analogue of the bonding charge
familiar from the theory of the covalent bond, although it is much
weaker and of different shape.

%%%%%%%%%%%%%%%%%%%%%%%%%%%%%%%%%%%%%%%%%%%%%%%%%%%%%%%%%%%%%%%%%%%%%%%%
\section{Conclusions}
\label{sec:conclusions}
%%%%%%%%%%%%%%%%%%%%%%%%%%%%%%%%%%%%%%%%%%%%%%%%%%%%%%%%%%%%%%%%%%%%%%%%

The long-range part of the Kohn-Sham exchange correlation potential
derived here allows the vdW-DF theory of Dion et al.\cite{dion04} to be
applied in a fully self-consistent manner.  In previous work the vdW-DF
functional had been applied as a post-GGA perturbation. It was argued
that this was a reasonable approximation, because one does not expect
that the rather weak and diffuse vdW interaction should substantially
change the electronic charge distribution.  The present work, through the
application of the fully self-consistent theory to a number of van der
Waals complexes, shows that this argument was correct. In all the
complexes studied, the predictions of the fully self-consistent theory
were nearly indistinguishable from those obtained via post-GGA
perturbation.  This was true near the equilibrium separation between the
fragments. However, inaccuracies in the perturbation method began to
become noticeable when the fragments were pushed together more closely
and the perturbation becomes larger.  The above conclusions
additionally validate the many previously performed
post-GGA perturbative calculations, and suggest that the more efficient
post-GGA perturbative method will remain an effective tool.  

Nevertheless, the availability of a self-consistent theory is important,
because it unleashes the possibility of applying a number of important
and standard DFT techniques whose availability depends on this
self-consistency.  The most obvious of these is the ability to calculate
internuclear forces via electrostatics. Indeed we show that the
Hellmann-Feynman forces are given accurately in vdW-DF, a feature that
will allow for efficient nuclear relaxation methods to be employed to
determine optimal geometric structures.

Finally, our applications to the geometries of isolated molecules  as
well as bulk silicon suggest that vdW-DF is really a general  density
functional. It  gives good results for systems or parts thereof  where
van der Waals forces are substantial, but does not spoil the results of
ordinary LDA-GGA type functionals  when van der Waals forces are
unimportant.

These studies have shown the strengths of the existing functional. However, areas for future improvements are also important.
One of these is the consistent small overestimation of the separation
between vdW bound fragments. This occurs in all the vdW complexes
considered here, as well as those considered in previous
work.\cite{aaron05,Thonhauser06,benz-graph,pe} The reason for this
persistent shift has been attributed to inadequacies of the exchange
functional,\cite{aaron05, Thonhauser06,pe} but a systematic study has
yet to be done. It seems likely that the source of this discrepancy can
be unambiguously identified and corrected in future work.

%%%%%%%%%%%%%%%%%%%%%%%%%%%%%%%%%%%%%%%%%%%%%%%%%%%%%%%%%%%%%%%%%%%%%%%%
\begin{acknowledgments}
We are grateful to Henrik Rydberg and Bengt Lundqvist, without whom
there would not have been this functional, to Maxime Dion for proving
its usefulness for real van der Waals complexes, and to Elsebeth
Schr{\"o}der, who led the development of methods for handling extended
systems. This work was supported by NSF Grant No.\ DMR-0456937. All
calculations were performed on the Rutgers high-performance
supercomputer facility, operated by the Center for Materials Theory of
the Department of Physics and Astronomy. PH thanks the Swedish Research
Council (VR).
\end{acknowledgments}
%%%%%%%%%%%%%%%%%%%%%%%%%%%%%%%%%%%%%%%%%%%%%%%%%%%%%%%%%%%%%%%%%%%%%%%%

%%%%%%%%%%%%%%%%%%%%%%%%%%%%%%%%%%%%%%%%%%%%%%%%%%%%%%%%%%%%%%%%%%%%%%%%
\begin{appendix}
%%%%%%%%%%%%%%%%%%%%%%%%%%%%%%%%%%%%%%%%%%%%%%%%%%%%%%%%%%%%%%%%%%%%%%%%

%%%%%%%%%%%%%%%%%%%%%%%%%%%%%%%%%%%%%%%%%%%%%%%%%%%%%%%%%%%%%%%%%%%%%%%%
\section{Details of the Evaluation of the Kernel}
\label{sec:details_a}
%%%%%%%%%%%%%%%%%%%%%%%%%%%%%%%%%%%%%%%%%%%%%%%%%%%%%%%%%%%%%%%%%%%%%%%%
%
The expression for the kernel $\phi$ can be written as\cite{dion04}
\begin{eqnarray}\label{equ:phiint}
\phi(d, d') &=& \frac{2}{\pi^2}\int_0^\infty\!a^2\,da\int_0^\infty\!
                b^2\,db\, W(a,b) \nonumber \\
&\times& T\left(\nu(a), \nu(b),\nu'(a),\nu'(b)\right)\;,
\end{eqnarray}
where the function $W(a,b)$ is defined as
\begin{eqnarray}\label{equ:Weval}
\lefteqn{W(a,b)= \frac{2}{a^3b^3}\Big[}\\
&& (3-a^2)b\cos b \sin a +(3-b^2)a \cos a \sin b \nonumber \\
&& {}+ (a^2+b^2-3)\sin a \sin b -3ab \cos a \cos b\Big]\;,\nonumber
\end{eqnarray}
and the function $T$ is given by
\begin{eqnarray}\label{equ:Teval}
T(w,x,y,z) &=& \frac{1}{2}\left[\frac{1}{w+x}+\frac{1}{y+z}\right]\\
&&\times\left[\frac{1}{(w+y)(x+z)} + \frac{1}{(w+z)(y+x)}\right]\;.
               \nonumber
\end{eqnarray}
Furthermore, we have used the definitions
\begin{subequations}
\label{equ:bothnudefs}
\begin{eqnarray}
\nu(u)&=&{u^2}/{2h(u/d)}  \;,\label{equ:nudef}\\
\nu'(u)&=&{u^2}/{2h(u/d')}\;,\label{equ:nuprimedef}
\end{eqnarray}
\end{subequations}
and
\begin{equation}
h(t) = 1-\exp{(- 4\pi t^2/9)}\;.
\end{equation}

Since $\phi$ according to \eq{equ:phiint} depends on $d$ only via $\nu$ 
and on $d'$ only via $\nu'$, it is straightforward to perform the
required derivatives indicated in \eqs{equ:phi} analytically, leaving
only the double integral over $a$ and $b$ in \eq{equ:phiint} to be done
numerically. The required partial derivatives of $T\left(\nu(a),
\nu(b),\nu'(a),\nu'(b)\right)$ in \eq{equ:phiint} are $T_{d}$, $T_{dd}$,
and $T_{dd'}$, where  $\nu$ and $\nu'$ depend implicitly on $d$ and $d'$
respectively according to \eqs{equ:bothnudefs}. Thus, $T_{d}$, $T_{dd}$,
and $T_{dd'}$ may be expressed in terms of partial  derivatives of
$T(w,x,y,z)$ in \eq{equ:Teval} with respect to $w$, $x$, $y$, $z$, and
the partial derivatives of $\nu$ and $\nu'$ in \eqs{equ:bothnudefs} with
respect to $d$ and $d'$. This procedure gives
\begin{subequations}
\label{equ:Tderivs}
\begin{eqnarray}
T_d     &=&	T_w\nu_d(a)	+ T_x\nu_d(b)\;,\label{Td}\\
T_{dd}  &=&	T_{ww}\nu_d^2(a) + T_w\nu_{dd}(a) + T_{xx}\nu_d^2(b)\nonumber\\
        &&  {}+T_x\nu_{dd}(b) + 2T_{wx}\nu_d(a)\nu_d(b)\;,\label{Tdd}\\
T_{dd'} &=& T_{wy}\nu'_{d'}(a)\nu_d(a) + T_{wz}\nu'_{d'}(b)\nu_d(a)\nonumber \\
        &&  {}+T_{xy}\nu'_{d'}(a)\nu_d(b) + T_{xz}\nu'_{d'}(b)\nu_d(b)\;.\label{Tddp}
\end{eqnarray}
\end{subequations}

The algebraic evaluation of the derivatives in \eqs{equ:Tderivs} is most
simply done with a symbolic algebra program. Then, altering
\eq{equ:phiint} by the replacement of $T$ with $dT_d$ using \eq{Td}
gives $\Phi_1$ (\eq{equ:phi1}). An analogous replacement using
$d^2T_{dd}$ gives $\Phi_2$ (\eq{equ:phi2}), while for $\Phi_3$
(\eq{equ:phi3}) the replacement should be made with
$T_d+dT_{dd}+d'T_{dd'}$.

%%%%%%%%%%%%%%%%%%%%%%%%%%%%%%%%%%%%%%%%%%%%%%%%%%%%%%%%%%%%%%%%%%%%%%%%
\section{Gradient correction}
\label{gradient}
%%%%%%%%%%%%%%%%%%%%%%%%%%%%%%%%%%%%%%%%%%%%%%%%%%%%%%%%%%%%%%%%%%%%%%%%

As discussed in Ref.~[\onlinecite{dion04}], the long range part of the
correlation energy $E_c^{\text{nl}}$ is calculated assuming a simple
single pole model for the necessary response function, with parameters
determined by sum rules plus the requirement that the correct energy
density should be obtained when it is applied locally, i.e., according
to LDA with the appropriate gradient correction.  In particular, the 
gradient terms that represent an (inappropriate) attempt to expand the
van der Waals interaction in a gradient series should be omitted.

The leading gradient contribution to the exchange-correlation energy
density may be written as
\begin{equation}
\label{egrad}
\varepsilon_\textrm{grad}=Z\frac{e^2 k_F}{12\pi}\left(\frac{\nabla n}{2k_Fn}\right)^2.
\end{equation}
This defines the quantity $Z$ according to the usage of Langreth and
Perdew\cite{LP1,LP2} and Langreth and Vosko.\cite{LV1} Rasolt and
Geldart\cite{RG1,RG2} define a $Z$ which is larger by an additive
constant of $4/3$, which would cause a corresponding change in
\eq{egrad}.  The sometimes confusing aspects of this notation have been
clarified more recently.\cite{LV2} The value of $Z$ was first inferred
by combining the calculations of Ma and Brueckner\cite{MaB} and
Sham,\cite{sham} giving the result $Z=1.1978$. The positive sign was
unexpected, and gave a correction to the LDA that went in the  wrong
direction, setting back the application of gradient corrections for
years.

The first step in the resolution of this puzzle came with Rasolt and
Geldart's breakdown of $Z$ according to the process involved, dividing
$Z=Z_{ab}+Z_c$, according to the processes depicted in Fig.~1a,b,c of
Ref.~[\onlinecite{RG1}] and Fig.~3a,b,c of Ref.~[\onlinecite{RG2}], and also
Fig.~4a,b,c of Ref.~[\onlinecite{LP2}] where the same results were
obtained by a different method. The inset c in each of these figures
shows what is known as the fluctuation diagram.  It is the cause of the
unphysical positive contribution to $Z$: one finds  $Z_{ab}=-0.8491$ and
$Z_c=2.0470$.  

It would be another decade, however, before the implications of the
fluctuation diagram would become fully apparent. This began with a
seminal paper by Maggs and Ashcroft,\cite{MaggsAshcroft} who showed that
the diagram was associated with the van der Waals interaction, followed
by further development,\cite{LV1,RapcewiczAchcroft} and
culminating\cite{Ylva} with the use of the fluctuation diagram to obtain
the London formula for the van der Waals interaction between two atoms. 
The reason for the failure of the gradient expansion was now apparent:
the calculation of $Z_c$ represented an attempt to expand the long-range
vdW interaction in powers of density gradients, an enterprise that was
in retrospect obviously doomed to fail.

Based on these facts, we write for use as the second term on the right
side of Eq.~(12) of Ref.~[\onlinecite{dion04}]
\begin{equation}
 \varepsilon_\textrm{grad}=Z_{ab}\frac{e^2 k_F}{12\pi}\left(\frac{\nabla n}{2k_Fn}\right)^2.
\label{egrad2}
\end{equation}
The van der Waals contribution is already included in another way, and
it would be double counting to add the gradient expansion to it in
addition. This is a matter of principle, but convenient indeed, because
it avoids the use of $Z_c$, which fails to be a valid approximation.
This explains the use of $Z_{ab}$ rather than $Z$ in \eq{q0}.

In principle, $Z_{ab}$ is a not a constant, but rather a function of
electronic density. There is no published data on this dependence. In
Fig.~2 of Ref.~[\onlinecite{LP1}] are shown two calculations that give the
density dependence of the full $Z$, one of which was obtained
earlier\cite{RG1,RG2} [$Z$ is equal to the ordinate of that figure times
a numerical constant]. When the calculations of Ref.~[\onlinecite{LP1}]
were made, the density dependence of $Z_{ab}$ and $Z_c$ were calculated
separately, but the significance of the individual quantities was not
known at that time, and individual results were not presented. However,
the common author of this paper and Ref.~[\onlinecite{LP1}], who did this
part of the calculation, attests that the principal density dependence
shown in the Fig.~2 of this reference comes from that of $Z_c$, and that
the density dependence of  $Z_{ab}$ was weak.  We thus feel comfortable
using the high density value of $Z_{ab}=-0.8491$.

%%%%%%%%%%%%%%%%%%%%%%%%%%%%%%%%%%%%%%%%%%%%%%%%%%%%%%%%%%%%%%%%%%%%%%%%
\section{Evaluation of the gradient}
\label{sec:details_b}
%%%%%%%%%%%%%%%%%%%%%%%%%%%%%%%%%%%%%%%%%%%%%%%%%%%%%%%%%%%%%%%%%%%%%%%%
%
In the following, we will focus on the term $\nabla\cdot[\dots]$ in
Eq.~(\ref{equ:v_xc_40}). It is straight forward to see that
\begin{eqnarray*}
\lefteqn{\nabla\cdot\Big[\phi_d(\rr,\rr')R_{\rr\rr'}
   \redgrad(\rr)\Big] = \big(\nabla\phi_d(\rr,\rr')\big)
   R_{\rr\rr'}\cdot\redgrad(\rr) + }\\
&& +\phi_d(\rr,\rr')\big(\nabla R_{\rr\rr'}\big)\cdot\redgrad(\rr)
   +\phi_d(\rr,\rr')R_{\rr\rr'}\nabla\cdot\redgrad(\rr)\;.
\end{eqnarray*}
The gradient of $R_{\rr\rr'}$ is the unit vector $\hat{\RR}_{\rr\rr'}$. 
For the gradient of $\phi_d(\rr,\rr')$ we abbreviate higher derivatives
of $\phi(d,d')$ in the same fashion as above,
\begin{equation*}
\frac{\partial^2\phi(d,d')}{\partial d\partial d} = \phi_{dd}(d,d')
\quad\text{and}\quad
\frac{\partial^2\phi(d,d')}{\partial d\partial d'} = \phi_{dd'}(d,d')\;.
\end{equation*}
Inserting the definition of $d(\rr,\rr')$ and $d'(\rr,\rr')$ from
Eq.~(\ref{equ:d_def}), the gradient of $\phi_d(\rr,\rr')$ can then
be written as
\begin{eqnarray}
\nabla\phi_d(\rr,\rr') &=& \phi_{dd}(\rr,\rr')\nabla d(\rr,\rr')
   + \phi_{dd'}(\rr,\rr')\nabla d'(\rr,\rr')\nonumber\\
&=& \phi_{dd}(\rr,\rr')\big(\hat{\RR}_{\rr\rr'}q_0(\rr) + 
     R_{\rr\rr'}\nabla q_0(\rr)\big)\nonumber\\
&& {}+ \phi_{dd'}(\rr,\rr')\hat{\RR}_{\rr\rr'}q_0(\rr')\;.
\end{eqnarray}
If we collect corresponding terms, the complete gradient
in Eq.~(\ref{equ:v_xc_40}) is then easily seen to be
\begin{eqnarray*}
\lefteqn{\nabla\cdot\Big[\phi_d(\rr,\rr')R_{\rr\rr'}
   \redgrad(\rr)\Big] = \hat{\RR}_{\rr\rr'}\cdot\redgrad(\rr)\;\times}\\
&& \Big[\phi_d(\rr,\rr') + \phi_{dd}(\rr,\rr')q_0(\rr)R_{\rr\rr'}
   +\phi_{dd'}(\rr,\rr')q_0(\rr')R_{\rr\rr'}\Big]\\
&& {} + \phi_{dd}(\rr,\rr')R_{\rr\rr'}^2\redgrad(\rr)\cdot\nabla q_0(\rr)
   + \phi_d(\rr,\rr')R_{\rr\rr'}\nabla\cdot\redgrad(\rr)\;.
\end{eqnarray*}
Expression for $\phi_d(d,d')$, $\phi_{dd}(d,d')$, and
$\phi_{dd'}(d,d')$ are discussed in Appendix~\ref{sec:details_a}.

%%%%%%%%%%%%%%%%%%%%%%%%%%%%%%%%%%%%%%%%%%%%%%%%%%%%%%%%%%%%%%%%%%%%%%%%
\end{appendix}
%%%%%%%%%%%%%%%%%%%%%%%%%%%%%%%%%%%%%%%%%%%%%%%%%%%%%%%%%%%%%%%%%%%%%%%%

%%%%%%%%%%%%%%%%%%%%%%%%%%%%%%%%%%%%%%%%%%%%%%%%%%%%%%%%%%%%%%%%%%%%%%%%

\end{document}